\documentstyle[aps,twocolumn,epsf,floats]{revtex}
\draft
\begin{document}
\wideabs{

\title{
   Energy spectra and photoluminescence of charged magneto-excitons}

\author{
   Arkadiusz W\'ojs*$\dagger$, John J. Quinn* and Pawel Hawrylak$\ddagger$}

\address{
   *Department of Physics, 
   University of Tennessee, Knoxville, Tennessee 37996, USA\\
   $\dagger$Institute of Physics, 
   Wroclaw University of Technology, Wroclaw 50-370, Poland\\
   $\ddagger$Institute for Microstructural Sciences, 
   National Research Council of Canada, Ottawa, Ontario K1A 0R6, Canada}

\maketitle

\begin{abstract}
Charged magneto-excitons $X^-$ in a dilute 2D electron gas in narrow 
and symmetric quantum wells are studied using exact diagonalization 
techniques. 
An excited triplet $X^-$ state with a binding energy of about 1~meV 
is found.
This state and the singlet are the two optically active states observed 
in photoluminescence (PL).
The interaction of $X^-$'s with electrons is shown to have short range, 
which effectively isolates bound $X^-$ states from a dilute $e$--$h$ 
plasma.
This results in the insensitivity of PL to the filling factor $\nu$.
For the ``dark'' triplet $X^-$ ground state, the oscillator strength 
decreases exponentially as a function of $\nu^{-1}$ which explains 
why it is not seen in PL.
\end{abstract}

\pacs{Keywords: 
   Charged magneto-exciton;
   Low-dimensional structure;
   Fractional quantum Hall effect}
}

Recent magneto-photoluminescence (PL) experiments\cite{hayne} showing 
recombination of charged excitons $X^-$ (two electrons bound to 
a valence hole) in narrow GaAs quantum wells appear to disagree 
completely with theoretical prediction\cite{whittaker}.
According to theory, the singlet (spin unpolarized) state $X^-_s$ is the 
$X^-$ ground state (GS) at low magnetic field, while the triplet $X^-_t$ 
is the GS at fields above 30~T.
In the PL experiments, the $X^-_s$ appears to be the GS for all magnetic 
fields, and $X^-_s$ and $X^-_t$ have comparable PL intensity.
Here we present results of numerical diagonalization of small systems, 
including effects of Landau level mixing and finite well widths.
We find that the energy of the lowest triplet state ($X^-_{td}$) behaves
exactly as predicted by previous calculations, but that its PL intensity 
is orders of magnitude smaller than those of the $X^-_s$ and an excited 
triplet state ($X^-_{tb}$).
We suggest that the triplet observed in PL is this bright triplet $X^-_{tb}$ 
whose energy is always higher than that of $X^-_s$.
The dark triplet $X^-_{td}$ is not observed in PL, and no disagreement
exists between theory and experiment.

The energy and PL spectra of the $X^-$ are calculated by exact numerical 
diagonalization of the two-electron--one-hole ($2e$--$1h$) Hamiltonian
\cite{x-fqhe,x-cf}.
In order to preserve the 2D translational symmetry of an infinite 
quantum well (QW) in a finite-size calculation, we use Haldane's
\cite{haldane} spherical geometry.
The magnetic field $B$ perpendicular to the surface of the sphere of
radius $R$ is due to a magnetic monopole placed in the center.
The monopole strength $2S$ is defined in the units of elementary flux
$\phi_0=hc/e$, so that $4\pi R^2B=2S\phi_0$, and the magnetic length 
is $\lambda=R/\sqrt{S}$.
The electron and hole states form degenerate angular momentum ($l$) 
shells or Landau levels (LL), and the lowest LL has $l=S$.
Our model applies to narrow and symmetric heterostructures, and the 
numerical results presented here are for the GaAs QW of width 11.5~nm.
For such a system, only the lowest QW subband need be included, 
and the cyclotron motion of both electrons and holes is well 
described in the effective-mass approximation, with the inter-subband 
coupling partially taken into account through the dependence of 
the hole cyclotron mass on $B$.
The finite (and different) widths of electron and hole quasi-2D 
layers are included through effective 2D interaction potentials.
The electron Zeeman energy depends on well width and $B$.
Five electron and hole LL's are used in the calculation, and the 
energies obtained for different values of $2S\le20$ are extrapolated 
to the limit of $S^{-1}=(\lambda/R)^2\rightarrow0$ (i.e. to the planar 
geometry), so that the finite-size and surface-curvature effects are 
eliminated.

The $2e$--$1h$ energy spectra (binding energy as a function of angular 
momentum $L$) calculated for $2S=20$ are shown in 
\begin{figure}[h]
\epsfxsize=3.40in
\epsffile{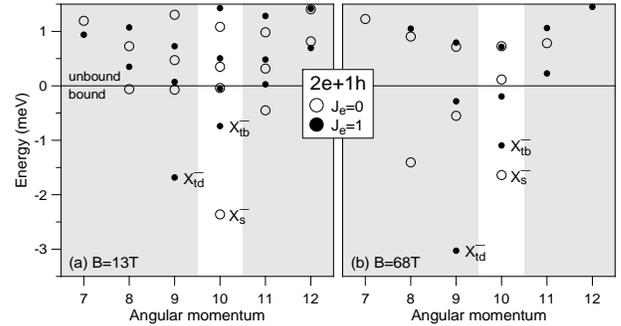}
\caption{
   The energy spectra (binding energy vs.\ angular momentum) of the 
   $2e$--$1h$ system on a Haldane sphere with the Landau level degeneracy 
   of $2S+1=21$.
   The parameters are appropriate for the 11.5~nm GaAs quantum well.
}
\label{fig1}
\end{figure}
Fig.~\ref{fig1}.
Open and full symbols mark singlet and triplet states ($J_e$ is the 
total electron spin), and each state with $L>0$ represents a degenerate 
$L$-multiplet.
Since the PL process (annihilation of an $e$--$h$ pair and emission of 
a photon) occurs with conservation of angular momentum, only states from 
the $L=S$ channel are radiative\cite{x-fqhe}.
Recombination of other, non-radiative states requires breaking rotational 
symmetry (e.g., by collisions with electrons).
This result is {\em independent} of the chosen spherical geometry and 
holds also for a planar QW, except that the definition of $L$ is different.

Three states marked in Fig.~\ref{fig1} are of particular importance:
$X^-_s$ and $X^-_{tb}$ are the only well bound radiative states, while 
$X^-_{td}$ has by far the lowest energy of all non-radiative states.
{\em The radiative triplet bound state $X^-_{tb}$ is identified for the 
first time.}
The binding energies of all three $X^-$ states are extrapolated to 
$\lambda/R\rightarrow0$ and plotted in 
\begin{figure}[t]
\epsfxsize=3.40in
\epsffile{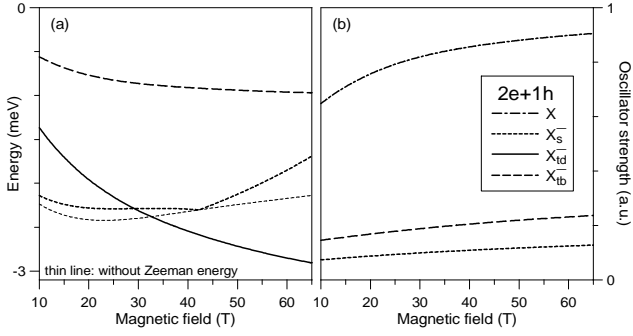}
\caption{
   The $X^-$ binding energies (a) and oscillator strengths (b) in the 
   11.5~nm GaAs quantum well plotted as a function of the magnetic field.
}
\label{fig2}
\end{figure}
Fig.~\ref{fig2}(a) as a function of $B$.
For the $X^-_s$, the binding energy differs from the PL energy (thin 
dotted line) by the Zeeman energy needed to flip one electron's spin,
and the cusp at $B\approx42$~T is due to the change of sign of the 
electron $g$-factor.
For the triplet states, the PL and binding energies are equal.
The energies of $X^-_s$ and $X^-_{td}$ behave as expected:
The binding of $X^-_s$ weakens at higher $B$ due to the ``hidden symmetry''
\cite{lerner}, which eventually leads to its unbinding in the infinite 
field limit\cite{macdonald}; the binding energy of $X^-_{td}$ changes as 
$e^2/\lambda\propto\sqrt{B}$; and the predicted\cite{whittaker} transition 
from the $X^-_s$ to the $X^-_{td}$ GS at $B\approx30$~T is confirmed.
The new $X^-_{tb}$ state remains an excited triplet state at all values 
of $B$, and its binding energy is smaller than that of $X^-_s$ by about 
1.5~meV.
The oscillator strengths $\tau^{-1}$ of a neutral exciton $X$ and the 
two radiative $X^-$ states are plotted in Fig.~\ref{fig2}(b).
In the $2e$--$1h$ spectrum, the well bound $X^-_s$ and $X^-_{tb}$ states 
share a considerable part of the total oscillator strength of one $X$,
with $\tau_{tb}^{-1}$ nearly twice larger than $\tau_s^{-1}$.

The comparison of calculated magnitude and magnetic field dependence 
of the $X^-$ binding energies with the experimental PL spectra
\cite{hayne,shields,finkelstein}, as well as high oscillator strength 
of the $X^-_{tb}$, lead to the conclusion that the three peaks (without 
counting the Zeeman splittings) observed in PL experiments are due to 
the recombination of $X$, $X^-_s$, and $X^-_{tb}$.
Due to the vanishing oscillator strength, the lowest triplet state 
$X^-_{td}$ found in earlier calculations\cite{whittaker,x-dot,palacios} 
remains undetected even at $B>30$~T, when it is expected to be the 
$X^-$ GS.
Only partial hole spin polarization at lower $B$ and its increase with
increasing $B$ can lead to an observed\cite{hayne} enhancement of the 
$X^-_{tb}$ PL intensity by up to a factor of two, while the intensity 
of the $X^-_s$ peak remains roughly unchanged.

The results in Figs.~\ref{fig1} and \ref{fig2} are quantitatively 
correct for narrow and symmetrically (or remotely) doped QW's.
In strongly asymmetric QW's or heterojunctions\cite{priest}, 
significant difference between electron and hole QW confinements 
increases the $e$--$h$ attraction compared to the $e$--$e$ repulsion, 
and the binding energies of all three $X^-$ states contain an 
uncompensated $e$--$h$ attraction which scales with $B$ like the 
exciton energy ($e^2/\lambda\propto\sqrt{B}$).
Nevertheless, our most important qualitative result remains valid 
for all structures: 
{\em The triplet $X^-$ state seen in PL is the bright excited triplet 
state $X^-_{tb}$ and not the lowest triplet state $X^-_{td}$.}

To understand why the $X^-_{td}$ state remains optically inactive 
even in the presence of collisions, the $e$--$X^-$ interactions must 
be studied in greater detail.
In 
\begin{figure}[t]
\epsfxsize=3.40in
\epsffile{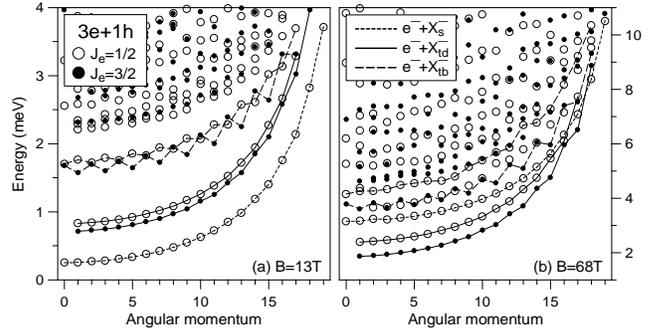}
\caption{
   The energy spectra (energy vs.\ angular momentum) of the $3e$--$1h$ 
   system on a Haldane sphere with the Landau level degeneracy 
   of $2S+1=21$.
   The parameters are appropriate for the 11.5~nm GaAs quantum well.
}
\label{fig3}
\end{figure}
Fig.~\ref{fig3} we plot the energy spectra of an $3e$--$1h$ system.
As in Fig.~\ref{fig1}, the energy is measured from the exciton energy 
and the open and filled circles mark multiplets with different $J_e$. 
In the low energy states, bound $X^-$ complexes interact with an electron 
through the effective pseudopotentials $V(L)$, defined as the dependence 
of pair interaction energy on pair angular momentum.
The pair angular momentum $L$ is related to the average $e$--$X^-$ 
separation $d$, and (on a sphere) larger $L$ corresponds to smaller $d$.
The allowed values of $J_e$ and $L$ can be understood by addition of
spins and angular momenta of an appropriate $X^-$ and an electron.
The total energy of an interacting pair is the sum of the $e$--$X^-$ 
repulsion energy $V(L)$ and the appropriate binding energy.
Because of incompatible energy scales, the $e$--$X^-$ scattering is 
nearly decoupled from internal $X^-$ excitations, and $V(L)$ is similar
for all pairs.
The relative position of $3e$--$1h$ energy bands corresponding to different 
$X^-$ complexes depends on the involved binding energy (and hence on $B$).

In narrow ($\le20$~nm) QW's, the $e$--$X^-$ pseudopotential retains 
the short-range character which results in the Laughlin correlations
\cite{laughlin}.
In a 2D electron gas, these correlations are responsible for the 
occurrence of the incompressible liquid states and the fractional 
quantum Hall effect.
Similar $e$--$X^-$ correlations in the $e$--$h$ plasma limit angular 
momentum (and energy) of $e$--$X^-$ collisions and, at low density 
and temperature, forbid an $X^-$ from getting close to an electron, 
effectively isolating it from the surrounding electron gas
\cite{x-fqhe,x-cf}.
This result depends critically on the short-range nature of $V(L)$, 
and thus on the QW thickness (in thicker QW's, high energy collisions 
occur even at low density).
The Laughlin correlations at the filling factor $\nu\le m^{-1}$ 
(where $\nu^{-1}$ is the number of magnetic flux quanta per electron) 
mean that all pair states with $L>2S-m$ are avoided\cite{parentage}.
This relates $\nu$ (i.e. density) to the maximum allowed $L$ (i.e. 
minimum distance) for an $e$--$X^-$ pair.

\begin{figure}[t]
\epsfxsize=3.40in
\epsffile{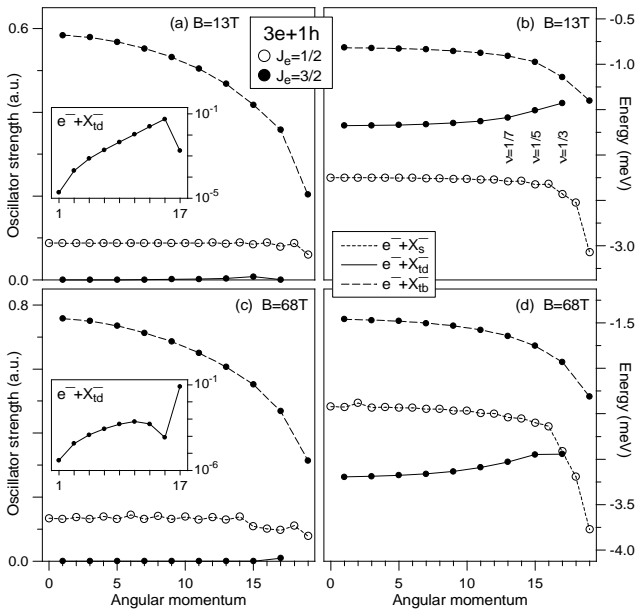}
\caption{
   The oscillator strengths (left) and recombination energies (right) 
   of an $X^-$ interacting with an electron on a Haldane sphere with 
   the Landau level degeneracy of $2S+1=21$, plotted as a function 
   of the $e$--$X^-$ pair angular momentum.
   The parameters are appropriate for the 11.5~nm GaAs quantum well.
}
\label{fig4}
\end{figure}

In Fig.~\ref{fig4} we plot the PL oscillator strength and energy 
(measured from the exciton energy) calculated for some of the $e$--$X^-$ 
states marked in 
Fig.~\ref{fig3}.
We assume that the Zeeman energy will polarize all electron spins prior 
to recombination, except for those two in the $X^-_s$, and concentrate 
on the following three initial configurations: $e$--$X^-_s$ with $J_e=
{1\over2}$, and $e$--$X^-_{tb}$ and $e$--$X^-_{td}$ with $J_e={3\over2}$.
For each of the three configurations, $\tau^{-1}$ and energy are plotted 
as a function of $L$ (i.e.\ of $\nu$).
The $e$--$X^-$ interactions have no significant effect on the PL 
oscillator strength and energy of an $X^-$ at small $L$ (i.e., at low 
density).
This justifies a simple picture of PL in dilute $e$--$h$ plasmas.
In this picture, recombination occurs from a single isolated bound complex 
and hence is virtually insensitive to $\nu$.
Quite surprisingly, the Laughlin correlations prevent increase of the 
$X^-_{td}$ oscillator strength $\tau^{-1}_{td}$ through collisions with 
other charges.
The $\tau^{-1}_{td}$ decreases exponentially (see insets in Fig.~\ref{fig4}) 
with decreasing $\nu$, and $\tau_{td}$ remains ten times longer than 
$\tau_s$ even at $\nu={1\over3}$.
This explains the absence of an $X^-_{td}$ peak even in those PL spectra 
\cite{hayne,shields,finkelstein,priest} showing strong recombination of 
a higher-energy triplet state $X^-_{tb}$.

\paragraph*{Acknowledgment.}
A.W. and J.J.Q. acknowledge partial support by the Materials Research 
Program of Basic Energy Sciences, US Dept.\ of Energy.

\end{document}